\documentstyle[prl,aps,twocolumn]{revtex}

\begin{document} 
\draft 
\title{$\hbar$ expansion for the periodic orbit quantization by harmonic
       inversion}
\author{J\"org Main,$^1$ Kirsten Weibert,$^1$ and G\"unter Wunner$^2$}
\address{$^1$Institut f\"ur Theoretische Physik I,
         Ruhr-Universit\"at Bochum, D-44780 Bochum, Germany}
\address{$^2$Institut f\"ur Theoretische Physik und Synergetik,
         Universit\"at Stuttgart, D-70550 Stuttgart, Germany}

%\date{Received: \hskip 8.0 truecm ~}
\date{\today}
\maketitle

\begin{abstract}
Semiclassical spectra beyond the Gutzwiller and Berry-Tabor approximation for
chaotic and regular systems, respectively, are obtained by harmonic inversion
of the $\hbar$ expansion of the periodic orbit signal.
The method is illustrated for the circle billiard, where the semiclassical 
error is reduced by one to several orders of magnitude with respect to the
lowest order approximation used previously.

\end{abstract}

\pacs{PACS numbers: 03.65.Sq, 05.45.+b}

Semiclassical spectra can be obtained for both regular and chaotic systems
in terms of the periodic orbits of the system.
For chaotic dynamics the semiclassical trace formula was derived by
Gutzwiller \cite{Gut67,Gut90}, and for integrable systems the Berry-Tabor 
formula \cite{Ber76} is well known to be precisely equivalent to the EBK 
torus quantization \cite{Ein17}.
However, the semiclassical trace formulae are exact only in exceptional
cases, e.g., the geodesic motion on the constant negative curvature surface.
In general they are just the leading order terms of an infinite series in
powers of the Planck constant and the accuracy of semiclassical quantization
is still an object of intense investigation \cite{Pro93,Boa94,Pri98}.
Methods for the calculation of the higher order periodic orbit 
contributions were developed in \cite{Gas93,Alo93,Vat96}.
However, the $\hbar$ expansion of the periodic orbit sum does not solve
the general problem of the construction of the analytic continuation of
the trace formula.
The semiclassical trace formula usually does not converge in the physically
interesting region even when only the leading order terms in $\hbar$ are 
considered, and special techniques are necessary to overcome the convergence 
problems \cite{Cvi89,Aur92,Ber92}.
Up to now the $\hbar$ expansion for periodic orbit quantization is restricted
to systems with known symbolic dynamics, like the three disk scattering
problem, where cycle expansion techniques can be applied \cite{Alo93,Vat96}.

Recently the {\em harmonic inversion} technique \cite{Wal95,Man97} was 
proposed as a universal method for periodic orbit quantization 
\cite{Mai97b,Mai98}, which allows the analytic continuation of the 
non-convergent periodic orbit sum to the region where the semiclassical 
eigenvalues and resonances are located.
The power of this method was demonstrated by its wide applicability to
open and bound systems with both regular and chaotic classical dynamics.
However, the method was restricted to the conventional lowest order $\hbar$ 
approximation of the periodic orbit sum, i.e., it cannot be applied
straightforwardly to the $\hbar$ expansion of the periodic orbit sum.
In this Paper we overcome these problems and extend the method of periodic 
orbit quantization by harmonic inversion to the analysis of the $\hbar$ 
expansion of the periodic orbit sum.
When applied to the circle billiard, as a first example, the accuracy of 
semiclassical eigenvalues is improved by at least one to several orders 
of magnitude.
The method can be applied to a large variety of systems, i.e., it is not
restricted to problems which can be solved with cycle expansion techniques.

As previously \cite{Mai97b} we consider systems with a scaling property, 
i.e., where the shape of periodic orbits does not depend on the scaling 
parameter, $w$, and the classical action $S_{\rm po}$ scales as 
\begin{equation}
 S_{\rm po} = w s_{\rm po} \; .
\label{S_po}
\end{equation}
The scaling parameter plays the role of an inverse effective Planck constant,
i.e., $w\equiv\hbar_{\rm eff}^{-1}$, and the $\hbar$ expansion of the
periodic orbit sum can therefore be written as a power series in $w^{-1}$.
The semiclassical spectrum is given by
\begin{equation}
 \varrho(w) = -{1\over\pi} \, {\rm Im} \, g(w) \; ,
\end{equation}
with
\begin{equation}
   g(w)
 = \sum_{n=0}^\infty g_n(w)
 = \sum_{n=0}^\infty {1\over w^{n}} \sum_{\rm po} 
   {\cal A}_{\rm po}^{(n)} e^{is_{\rm po}w}
\label{g_sc}
\end{equation}
the fluctuating part of the semiclassical response function.
The ${\cal A}_{\rm po}^{(n)}$ are the complex amplitudes of the $n^{\rm th}$
order periodic orbit contributions including phase information from the 
Maslov indices.
Usually the zeroth order contributions ${\cal A}_{\rm po}^{(0)}$ are 
considered only.
The Fourier transform of the principal periodic orbit sum
\begin{eqnarray}
     C_0(s)
 &=& {1\over 2\pi}\int_{-\infty}^{+\infty} g_0(w) e^{-isw}dw
     \nonumber \\
 &=& \sum_{\rm po}{\cal A}_{\rm po}^{(0)} \delta(s-s_{\rm po})
\label{C0_sc}
\end{eqnarray}
is adjusted by application of the {\em harmonic inversion} technique
\cite{Mai97b,Mai98} to the functional form of the exact quantum expression
\begin{eqnarray}
     C(s)
 &=& {1\over 2\pi}\int_{-\infty}^{+\infty} \sum_k{d_k\over w-w_k+i\epsilon}
     e^{-iws}dw \nonumber \\
 &=& -i\sum_k d_k e^{-iw_ks} \; ,
\label{C_qm1}
\end{eqnarray}
with $\{w_k,d_k\}$ the eigenvalues and multiplicities.
The frequencies $w_{k,0}$ obtained by harmonic inversion of Eq.\ \ref{C0_sc} 
are the zeroth order $\hbar$ approximation to the semiclassical eigenvalues.
We will now demonstrate how the higher order correction terms to the 
semiclassical eigenvalues can be extracted from the periodic orbit sum 
(\ref{g_sc}).
We first remark that the asymptotic expansion (\ref{g_sc}) of the 
semiclassical response function suffers, for $n\ge 1$, from the singularities 
at $w=0$, and it is therefore not appropriate to harmonically invert the 
Fourier transform of (\ref{g_sc}), although the Fourier transform formally 
exists.
This means that the method of periodic orbit quantization by harmonic 
inversion cannot straightforwardly be extended to the $\hbar$ expansion of
the periodic orbit sum.
Instead we will calculate the correction terms to the semiclassical 
eigenvalues separately, order by order, as described in the following.

Let us assume that the $(n-1)^{\rm st}$ order approximations $w_{k,n-1}$ to
the semiclassical eigenvalues are already obtained and the $w_{k,n}$
are to be calculated.
The difference between the two subsequent approximations to the quantum
mechanical response function reads
\begin{eqnarray}
     g_{n}(w)
 &=& \sum_k \left({d_k\over w-w_{k,n}+i\epsilon}
          - {d_k\over w-w_{k,n-1}+i\epsilon}\right)
     \nonumber \\
 &\approx& \sum_k{d_k\Delta w_{k,n}\over (w-\bar w_{k,n}+i\epsilon)^2} \; ,
\label{g_n}
\end{eqnarray}
with $\bar w_{k,n}=(w_{k,n}+w_{k,n-1})/2$ and 
$\Delta w_{k,n}=w_{k,n}-w_{k,n-1}$.
Integration of Eq.\ \ref{g_n} and multiplication by $w^n$ yields
\begin{equation}
 {\cal G}_{n}(w) = w^n \int g_{n}(w)dw
 = \sum_k {-d_kw^n\Delta w_{k,n}\over w-\bar w_{k,n}+i\epsilon} \; ,
\label{g_int_qm}
\end{equation}
which has the functional form of a quantum mechanical response function but
with residues proportional to the $n^{\rm th}$ order corrections 
$\Delta w_{k,n}$ to the semiclassical eigenvalues.
The semiclassical approximation to Eq.\ \ref{g_int_qm} is obtained from 
the term $g_{n}(w)$ in the periodic orbit sum (\ref{g_sc}) by integration 
and multiplication by $w^n$, i.e.\
\begin{eqnarray}
     {\cal G}_{n}(w)
 &=& w^n \int g_{n}(w)dw \nonumber \\
 &=& -i\sum_{\rm po} {1\over s_{\rm po}}
     {\cal A}_{\rm po}^{(n)} e^{iws_{\rm po}}
     + {\cal O}\left(1\over w\right)  \; .
\label{g_int_sc}
\end{eqnarray}
We can now Fourier transform both Eqs.\ \ref{g_int_qm} and \ref{g_int_sc},
and obtain ($n\ge 1$)
\begin{eqnarray}
     C_{n}(s)
 &\equiv& {1\over 2\pi}\int_{-\infty}^{+\infty}{\cal G}_{n}(w)e^{-iws}dw
     \nonumber  \\
\label{C_qm}
 &=& i\sum_k d_k (w_k)^n\Delta w_{k,n} e^{-iw_ks} \\
\label{C_sc}
 &\stackrel{\rm h.i.}{=}&
     -i\sum_{\rm po}{1\over s_{\rm po}}{\cal A}_{\rm po}^{(n)}
     \delta(s-s_{\rm po})  \; .
\end{eqnarray}
Eqs.\ \ref{C_qm} and \ref{C_sc} are the main result of this Paper.
They imply that the $\hbar$ expansion of the semiclassical eigenvalues can 
be obtained, order by order, by harmonic inversion (h.i.) of the periodic
orbit signal in Eq.\ \ref{C_sc} to the functional form of Eq.\ \ref{C_qm}.
The frequencies of the periodic orbit signal (\ref{C_sc}) are the 
semiclassical eigenvalues $w_k$.
Note that the accuracy of the semiclassical eigenvalues does not necessarily
increase with increasing order $n$.
We indicate this in Eq.\ \ref{C_qm} by omitting the index $n$ at the 
eigenvalues $w_k$.
The corrections $\Delta w_{k,n}$ to the eigenvalues are obtained from
the {\em amplitudes}, $d_k(w_k)^n\Delta w_{k,n}$, of the periodic orbit
signal.

The method requires as input the periodic orbits of the classical system 
up to a maximum period (scaled action), $s_{\rm max}$, determined by the 
average density of states \cite{Mai97b,Mai98}.
The amplitudes ${\cal A}_{\rm po}^{(0)}$ are obtained from Gutzwiller's
trace formula \cite{Gut67,Gut90} and the Berry-Tabor formula \cite{Ber76}
for chaotic and regular systems, respectively.
For the next order correction ${\cal A}_{\rm po}^{(1)}$ explicit formulae
were derived by Gaspard and Alonso for chaotic systems with smooth potentials
\cite{Gas93} and in Refs.\ \cite{Alo93,Vat96} for billiards.
With appropriate modifications \cite{Wei98} the formulae can be used for 
regular systems as well.

We now demonstrate the $\hbar$ expansion of the periodic orbit sum for 
the example of the circle billiard.
We choose this system mainly for the sake of simplicity, since all the periodic
orbits and the relevant physical quantities can be obtained analytically.
It will be evident that the procedure works equally well with more complex
systems where periodic orbits have to be searched numerically.
Furthermore this system has served recently as a showpiece example for solving
the fundamental problem of reducing the number of orbits required for 
periodic orbit quantization \cite{Mai98b}.
The exact quantum mechanical eigenvalues $E=\hbar^2k^2/2M$ of the circle
billiard are given as zeros of Bessel functions $J_{|m|}(kR)=0$, where $m$ 
is the angular momentum quantum number and $R$, the radius of the circle.
In the following we choose $R=1$.
The lowest order semiclassical eigenvalues can be obtained by an EBK torus 
quantization resulting in the quantization condition \cite{Pro93}
\begin{equation}
 kR\sqrt{1-(m/kR)^2} - |m|\arccos{|m|\over kR} = \pi\left(n+{3\over 4}\right)
\label{EBK}
\end{equation}
with $m=0,\pm 1,\pm 2,\dots$ being the angular momentum quantum number and 
$n=0,1,2,\dots$ the radial quantum number.
States with angular momentum quantum number $m\ne 0$ are twofold degenerate
($d_k=2$).

For billiard systems the scaling parameter is the absolute value of the
wave vector, $w\equiv k=|{\bf p}|/\hbar$, and the action is proportional to 
the length of the orbit, $S_{\rm po}=\hbar k\ell_{\rm po}$.
The periodic orbits of the circle billiard are those orbits for which the 
angle between two bounces is a rational multiple of $2\pi$, i.e., the periods 
$\ell_{\rm po}$ are obtained from the condition
\begin{equation}
 \ell_{\rm po} = 2m_r \sin \gamma \; ,
\end{equation}
with $\gamma\equiv\pi m_\phi/m_r$, $m_\phi=1,2,\dots$ the number of turns 
of the orbit around the origin, and $m_r=2m_\phi,2m_\phi+1,\dots$ the number 
of reflections at the boundary of the circle.
Periodic orbits with $m_r\ne 2m_\phi$ can be traversed in two directions
and thus have multiplicity 2.
As mentioned before the calculation of the zeroth order amplitudes 
${\cal A}_{\rm po}^{(0)}$ in Eq.\ \ref{g_sc} depends on whether the 
classical dynamics is regular or chaotic.
For the circle billiard with regular dynamics we start from the Berry-Tabor 
formula \cite{Ber76} and obtain
\begin{equation}
   {\cal A}_{\rm po}^{(0)}
 = \sqrt{\pi\over 2}{\ell_{\rm po}^{3/2}\over m_r^2}
   e^{-i({\pi\over 2}\mu_{\rm po}+{\pi\over 4})}  \; ,
\label{A0_circ}
\end{equation}
where $\mu_{\rm po}=3m_r$ is the Maslov index.
[Note that the factor $w^{-n}$ in Eq.\ \ref{g_sc} must be replaced 
by $w^{-(n-1/2)}$ for the regular circle billiard.]
For the calculation of the first order periodic orbit contribution $g_1(w)$
in Eq.\ \ref{g_sc} we adopt the method of Alonso and Gaspard \cite{Alo93}
and obtain the first order periodic orbit amplitudes
\begin{equation}
   {\cal A}_{\rm po}^{(1)}
 = \sqrt{\pi m_r} \, {5-2\sin^2\gamma\over 6\sqrt{\sin^3\gamma}}
   e^{-i({\pi\over 2}\mu_{\rm po}-{\pi\over 4})}  \; .
\label{A1_circ}
\end{equation}
A detailed derivation of Eq.\ \ref{A1_circ} will be given elsewhere
\cite{Wei98}.
With the periodic orbit amplitudes, Eqs.\ \ref{A0_circ} and \ref{A1_circ}
at hand we have all the ingredients necessary for the harmonic inversion of 
the zeroth and first order periodic orbit signal.
For the technical details of the harmonic inversion technique see Refs.\ 
\cite{Wal95,Man97,Mai98}.
We considered periodic orbits up to maximum length $\ell_{\rm max}=200$,
which was sufficient to resolve the low lying states, despite a few near 
degeneracies.
The zeroth order semiclassical approximations $k^{(0)}$ to the eigenvalues 
are obtained by harmonic inversion of the signal $C_0(s)$ (Eq.\ \ref{C0_sc}),
and are presented in Table \ref{table1}.
They agree (despite the near degenerate states at $k\approx 11.05$ marked by 
asterisks) within the numerical accuracy with the results of the torus 
quantization, Eq.\ \ref{EBK} (see eigenvalues $k^{\rm EBK}$ in Table 
\ref{table1}).
However, the semiclassical eigenvalues deviate significantly, especially
for states with low radial quantum numbers $n$, from the exact quantum
mechanical eigenvalues $k^{\rm ex}$ in Table \ref{table1}.

\begin{figure}[t]
%\vspace{19.5cm}
%\special{psfile=FIG1.ps voffset=-45 hoffset=-25 vscale=78 hscale=78}
\vspace{5.4cm}
\includegraphics{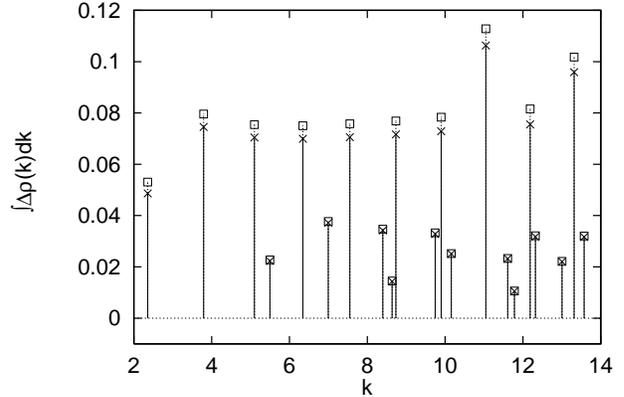}
\caption{\label{fig1} 
Integrated difference of the density of states, $\int\Delta\varrho(k)dk$,
for the circle billiard with radius $R=1$. 
Crosses: $\Delta\varrho(k)=\varrho^{\rm ex}(k)-\varrho^{\rm EBK}(k)$.
Squares: $\Delta\varrho(k)=\varrho^{\rm (1)}(k)-\varrho^{\rm (0)}(k)$
obtained from the $\hbar$ expansion of the periodic orbit signal.
}
\end{figure}

The first order corrections to the semiclassical eigenvalues $k^{(0)}$
are obtained by harmonic inversion of the periodic orbit signal
$C_1(s)$ (Eq.\ \ref{C_sc}).
The resulting spectrum, i.e., the integrated differences of the density 
of states $\int\Delta\varrho(k)dk$ are shown in Fig.\ \ref{fig1}.
The squares mark the spectrum for 
$\Delta\varrho(k)=\varrho^{\rm (1)}(k)-\varrho^{\rm (0)}(k)$
obtained from the harmonic inversion of the signal $C_1(s)$.
For comparison the crosses present the same spectrum but for the difference 
$\Delta\varrho(k)=\varrho^{\rm ex}(k)-\varrho^{\rm EBK}(k)$ between the 
exact quantum mechanical and the EBK-spectrum.
The deviations between the peak heights exhibit the contributions
\newpage
\phantom{}
\noindent
of terms of the $\hbar$ expansion series beyond the first order 
approximation.

\begin{figure}[t]
%\vspace{19.5cm}
%\special{psfile=FIG2.ps voffset=-45 hoffset=-25 vscale=78 hscale=78}
\vspace{5.6cm}
\includegraphics{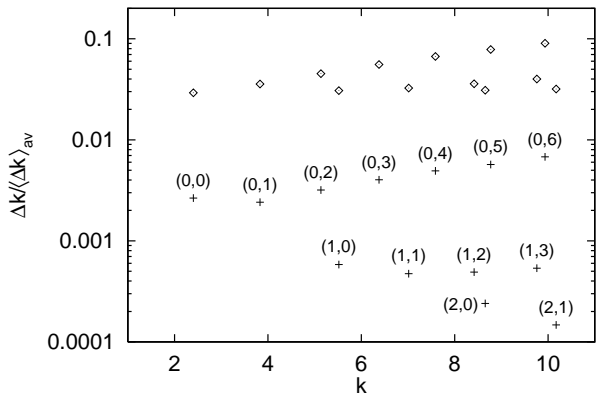}
\caption{\label{fig2} 
Semiclassical error $|k^{\rm (0)}-k^{\rm ex}|$ (diamonds) and 
$|k^{\rm (1)}-k^{\rm ex}|$ (crosses) in units of the average level spacing
$\langle\Delta k\rangle_{\rm av}\approx 4/k$.
States are labeled by quantum numbers $(n,m)$.
}
\end{figure}

The peak heights of the levels in Fig.\ \ref{fig1} (solid lines and crosses)
are, up to a multiplicity factor for the degenerate states, the shifts 
$\Delta k$ between the zeroth and first order semiclassical approximations 
to the eigenvalues $k$.
The
%
%\newpage
\begin{table}[t]
\caption{\label{table1}
The 20 lowest eigenstates of the circle billiard with radius $R=1$.
$n,m$: Radial and angular momentum quantum numbers;
$k^{\rm EBK}$: Results from EBK-quantization;
$k^{\rm (0)}$: Eigenvalues obtained by harmonic inversion of the periodic
orbit signal without $\hbar$ corrections 
(nearly degenerate states marked by asterisks are not fully resolved);
$k^{\rm (1)}$: Eigenvalues obtained by harmonic inversion of the periodic
orbit signal including $\hbar$ correction;
$k^{\rm ex}$: Exact eigenvalues, i.e., zeros of the Bessel functions 
$J_m(kR)=0$.}
\smallskip
\begin{center}
\begin{tabular}[t]{r|r|r|r@{.}l|r@{.}l|r}
  \multicolumn{1}{c|}{$n$} &
  \multicolumn{1}{c|}{$m$} &
  \multicolumn{1}{c|}{$k^{\rm EBK}$} &
  \multicolumn{2}{c|}{$k^{\rm (0)}$} &
  \multicolumn{2}{c|}{$k^{\rm (1)}$} &
  \multicolumn{1}{c}{$k^{\rm ex}$} \\ 
\hline
   0 &  0 &   2.356194 &    2&356187 &    2&409239 &   2.404826 \\
   0 &  1 &   3.794440 &    3&794430 &    3&834226 &   3.831706 \\
   0 &  2 &   5.100386 &    5&100379 &    5&138108 &   5.135622 \\
   1 &  0 &   5.497787 &    5&497782 &    5&520501 &   5.520078 \\
   0 &  3 &   6.345186 &    6&345180 &    6&382687 &   6.380162 \\
   1 &  1 &   6.997002 &    6&996999 &    7&015857 &   7.015587 \\
   0 &  4 &   7.553060 &    7&553053 &    7&590944 &   7.588342 \\
   1 &  2 &   8.400144 &    8&400140 &    8&417477 &   8.417244 \\
   2 &  0 &   8.639380 &    8&639370 &    8&653839 &   8.653728 \\
   0 &  5 &   8.735670 &    8&735652 &    8&774088 &   8.771484 \\
   1 &  3 &   9.744628 &    9&744619 &    9&761243 &   9.761023 \\
   0 &  6 &   9.899671 &    9&899663 &    9&938844 &   9.936110 \\
   2 &  1 &  10.160928 &   10&160925 &   10&173526 &  10.173468 \\
   1 &  4 &  11.048664 &  11&048966$^\ast$&  11&077169$^\ast$&  11.064709 \\
   0 &  7 &  11.049268 &  11&048966$^\ast$&  11&077169$^\ast$&  11.086370 \\
   2 &  2 &  11.608251 &   11&608248 &   11&619883 &  11.619841 \\
   3 &  0 &  11.780972 &   11&780968 &   11&791546 &  11.791534 \\
   0 &  8 &  12.187316 &   12&187318 &   12&228099 &  12.225092 \\
   1 &  5 &  12.322723 &   12&322717 &   12&338791 &  12.338604 \\
   2 &  3 &  13.004166 &   13&004163 &   13&015235 &  13.015201 \\
\end{tabular}
\end{center}
\end{table}
\noindent
first order eigenvalues $k^{(1)}=k^{(0)}+\Delta k$ are presented
in Table \ref{table1}, and are in excellent agreement with the exact 
eigenvalues $k^{\rm ex}$.
An appropriate measure for the accuracy of semiclassical eigenvalues is 
the deviation from the exact quantum eigenvalues in units of the average
level spacings, $\langle\Delta k\rangle_{\rm av}=1/\bar\varrho(k)$.
Fig.\ \ref{fig2} presents the semiclassical error in units of the average 
level spacings $\langle\Delta k\rangle_{\rm av}\approx 4/k$ for the zeroth
order (diamonds) and first order (crosses) approximations to the eigenvalues.
States are labeled by the radial and angular momentum quantum numbers $(n,m)$.
In the zeroth order approximation the semiclassical error for the low lying 
states is about 3 to 10 percent of the mean level spacing.
This error is reduced in the first order approximation by at least one order
of magnitude for the least semiclassical states with radial quantum number
$n=0$.
The accuracy of states with $n\ge 1$ is improved by two or more orders of 
magnitude.

As mentioned above the general technique developed in this Paper is not
restricted to the circle billiard but can in general be applied to the
whole variety of systems which can be quantized semiclassically by
harmonic inversion of the periodic orbit sum.
While for the circle billiard the periodic orbit parameters can be
calculated analytically the orbits must be obtained from a numerical
periodic orbit search in general.
However, no additional periodic orbits need to be searched for the $\hbar$
expansion of the periodic orbit sum, i.e., it is sufficient to calculate 
the amplitudes in Eq.\ \ref{g_sc} for the given set of orbits as described 
in Refs.\ \cite{Gas93,Alo93,Vat96}.

In conclusion, we have demonstrated that semiclassical spectra beyond the
Gutzwiller and Berry-Tabor approximation can be obtained by harmonic inversion
of the $\hbar$ expansion of the periodic orbit signal.
For the circle billiard, as a first example, the semiclassical error is 
reduced by at least one to several orders of magnitude by just including 
the lowest order periodic orbit correction terms.
The method proposed in this Paper opens the way to the calculation of high 
precision semiclassical eigenvalues directly from periodic orbit data for 
both regular and chaotic systems.
It is not restricted to bound systems but can be applied to open systems 
as well.

\bigskip
This work was supported by the Deutsche For\-schungs\-ge\-mein\-schaft 
(Sonder\-for\-schungs\-be\-reich No.\ 237).

\end{document}